\documentstyle[12pt,amssymb]{article}
\newcommand{\beq}{\begin{equation}}
\newcommand{\eeq}{\end{equation}}
\newcommand{\ba}{\begin{array}}
\newcommand{\ea}{\end{array}}
\newcommand{\bea}{\begin{eqnarray}}
\newcommand{\eea}{\end{eqnarray}}
\newcommand{\bean}{\begin{eqnarray*}}
\newcommand{\eean}{\end{eqnarray*}}

\newtheorem{theorem}{Theorem}[section]
\newtheorem{prop}[theorem]{Proposition}
\newtheorem{lem}[theorem]{Lemma}

\newtheorem{remark}[theorem]{Remark}
\newenvironment{rem}{\begin{remark} \rm}{\end{remark}}

\newcommand{\CN}{{\cal N}}

\makeatletter
\@addtoreset{equation}{section}

\makeatother

\newcommand{\cmp}[3]{Comm. Math. Phys. {\bf #1} (#2), #3}

\newcommand{\lanl}[1]{LANL preprint, hep-th/#1}
\newcommand{\jmp}[3]{Jour. Math. Phys. {\bf #1} (#2), #3}
\newcommand{\rref}[1]{(\ref{#1})} 

\newcommand{\del}{{\partial}}

\def\Fdb{{Fa\`a di Bruno}}

\def\dpt#1#2{\frac{\partial #1}{\partial t_{#2}}}

\def\H#1{H^{(#1)}}
\def\W#1{W^{(#1)}}

\def\h#1{h^{(#1)}}

\def\pf{\noindent {\bf Proof.}\ }
\def\epf{\hfill$\square$\medskip}

\begin{document}
\baselineskip=20pt
\begin{titlepage}
\begin{flushright}
Ref. SISSA 37/97/FM
\end{flushright}
\vspace{1.5truecm}
\begin{center}
{\huge Krichever Maps, Fa\`a di Bruno Polynomials,
\\ \vspace{.3truecm} and Cohomology in KP Theory}
\end{center}
\vspace{0.8truecm}
\begin{center}
{\large
Gregorio Falqui, Cesare Reina, \\ and Alessandro Zampa}
\par
\vspace{1.truecm}
SISSA/ISAS, Via Beirut 2/4, I-34014 Trieste, Italy\\
\vspace{.5truecm}
E--mail: falqui@sissa.it, reina@sissa.it, zampa@sissa.it\\
\end{center}
\vspace{1.5truecm}
\abstract{\noindent We study the geometrical meaning of the Fa\`a di
Bruno polynomials in the context of KP theory. They provide a
basis in a subspace $W$
of the universal Grassmannian associated to the KP hierarchy.
When $W$ comes from geometrical data via the Krichever
map, the Fa\`a di Bruno recursion relation turns out to be
the cocycle condition for (the Welters hypercohomology group describing) the
deformations of the dynamical line bundle on the spectral curve together with
the meromorphic sections which give rise to the Krichever map.
Starting from this, one sees that the whole KP
hierarchy has a similar cohomological meaning.}

\vspace{3.truecm}
Work partially supported by  M.U.R.S.T. and by
the G.N.F.M./C.N.R
\end{titlepage}
\setcounter{footnote}{0}
\section{Introduction}\label{introd}
The aim of this paper is to bridge between the algebro--geometrical
setting of KP theory ~\cite{DKN, Dub, Kr}, whose basic tools
are the Baker-Akhiezer function $\psi$ and the Hirota $\tau$ function,
and the construction based on the generating function
$h(z)=z+\sum_{i\ge 1} h_i z^{-i}$
of the hamiltonian densities and its associated
\Fdb\ polynomials.
Although dating back to the very
beginning of the modern theory of soliton
equations~\cite{MGK68,Ch78,Wil81},
up to now the second approach has been applied mostly
to the algebraic study of the KP theory and its reductions
(see, e.g.,~\cite{Ar95,DBook,Ma78}).
It has been reconsidered more
recently in~\cite{cfmp,mpz}
in the framework of the bihamiltonian theory of
integrable systems.
More precisely, the motivations of our work are the following:\\
a) It is well known~\cite{SS,sw}
that one can associate to any formal Baker-Akhiezer function $\psi$
of the KP theory a moving point $\W{\psi}$ in
the universal Grassmannian $Gr$ and that
the KP flows linearize there.
Another way of getting this result
algebraically~\cite{cfmp,fmp} is
by means of the Fa\`a di Bruno polynomials $h^{(k)}(z,x)$ recursively
defined by eq. \rref{recrel} below. These give rise to a basis in (another)
space $\W{h}\in Gr$. The linear flows for $\W{\psi}$
correspond to slightly more general Riccati type
equations for $\W{h}$.\\
b) For algebraic geometrical solutions, KP can be linearized on the
Jacobian of the spectral curve $C$ as well,
the link between the two linearizations
being the Krichever map~\cite{Kr,sw}. This map
associates a point $W\in Gr$ to the generic datum of a genus $g$ curve
$C$, a point $p\in C$, a local coordinate $z^{-1}$ vanishing at $p$,
a line bundle ${\cal L}_D$ of
degree $g$ and a local trivialization $\phi_0$ of ${\cal L}_D$ in a
neighborhood $U_0$ of $p$.

It is then
natural to ask for the geometrical meaning of the Fa\`a di Bruno basis.
The key observations to answer this question are the following:

\noindent 1) Generically, ${\cal L}_D$ has a unique (up to
homotheties) holomorphic section $\sigma_D$, with divisor $D$. If $p\in C$
is not contained in $D$, $\sigma_D$ does not vanish at
$p$ and gives a trivialization of ${\cal L}$ in a neighborhood of $p$.

\noindent 2) From the cohomology sequence
$$0\rightarrow H^0(C,{\cal O})\rightarrow H^0(C,{\cal O}(p))
{\buildrel{\Psi_4}\over{\rightarrow}}
{\Bbb C}\rightarrow H^1(C,{\cal O})\rightarrow H^1(C,{\cal
O}(p))\rightarrow 0$$
associated to the injection
${\cal O}\rightarrow {\cal O}(p)$, we get that $\Psi_4$ is an injection.
We denote by $[\alpha ]:=\Psi_4 (z)$ the image if the
generator $z$ of $H^0(C,{\cal O}(p)/{\cal O})$. So we have a
distinguished vector field tangent to $Pic_g(C)$.

Summing up, {\em the generic datum of $(C,p,z,{\cal L}_D)$ gives us a
trivialization
$\phi_0$ of ${\cal L}_D$ on $U_0\ni p$, i.e. a set of Krichever data,
together with infinitesimal deformations of ${\cal L}_D$,
corresponding to representatives $\alpha $ of $[\alpha]$}.

\noindent 3) We can then use the data above to deform ${\cal
L}_D$ together with
$\sigma_D$ along $\alpha$. The cocycle condition for such an
infinitesimal deformation (see eq.~\rref{newsect}) looks like the
Fa\`a di Bruno relation (see eq.~\rref{recrel}
but misses the $x$ dependence of the
Fa\`a di Bruno polynomials. To accommodate this dependence, we exploit the
existence of a universal family of line bundles over $C$
with a section and prove that one can choose $\alpha$ in such a way that
the Fa\`a di Bruno relation is actually the cocycle condition for the
infinitesimal deformation of the elements of this family.

That the KP hierarchy was related to cohomology has been
known since~\cite{Mu}.
This set up corresponds to the projection on $H^1(C,{\cal O})$ of
the cocycle conditions above, which control the infinitesimal
deformations of ${\cal L}_D$.
The novelty of our result is that, by considering also the
deformations of the sections of ${\cal L}_D$ which give rise to the
Krichever map, one gets as cocycle conditions the equations
of the KP hierarchy as a dynamical system on $Gr$~\cite{SS}.

Section~\ref{sec2}
quickly describes the appearance of
\Fdb\ polynomials in KP theory and the associated map
to $Gr$. The proofs of the main results are collected in Section~\ref{sec3}.
Section~\ref{sec4} is devoted to the explicit description
of the case of elliptic curves. Some notions
of deformation theory are recalled in Appendix~\ref{app}.
In the sequel, we will use the notations of [11] while dealing with
curves and their Jacobians. In particular we refer to [11], ch. 2.7 for the
results recalled in Section 3.

\section{\Fdb\ polynomials, the KP theory and the Grassmannian}\label{sec2}

The KP hierarchy is an isospectral deformation
of a monic operator of degree $1$
in the ring of pseudodifferential operators on the circle $S^1$.
Given such an operator
\begin{equation}\label{5.1}
Q=\del-\sum_{l\ge 1} q_i \del^{-i},
\end{equation}
one defines the associated linear problem for the Baker--Akhiezer
function $\psi$:
\begin{equation}\label{linpb}
\left\{
\begin{array}{rl}
Q \psi & = z \psi \\
\dpt{}{n} \psi &= Q^n_+ \psi,\end{array}\right.
\end{equation}
where $Q^n_+=Q^n-Q^n_-$ is the differential part of $Q^n$.
The KP hierarchy (see, e.g.,~\cite{DJKM})
is the set of compatibility conditions
of the linear system above,
i.e. the hierarchy of Lax equations
\begin{equation}
\begin{array}{rl}
\dpt{}{n}Q &= [Q^n_+,Q]\\
\dpt{Q^m_+}{n} -\dpt{Q^n_+}{m}&= [Q^m_+,Q^n_+].\end{array}
\label{djkm}
\end{equation}
A formally equivalent description starts with a monic Laurent series $h$
of the form $h=z+\sum_{l\ge 1 } h_l(x) z^{-l}$
with coefficients in the space of smooth functions on the circle,
$x$ being a coordinate on $S^1$.
Its \Fdb\ iterates $\h{k}$ are defined by the recurrence relation
\begin{equation}\label{recrel}
\h{k+1}=\del_x\h{k}+h\h{k},\quad \h{0}:= 1.
\end{equation}
Since $\h{k}$ has a Laurent expansion
$\h{k}=z^k+O(z^{k-2})$,
the equation $z=h-\sum_{j\geq 1}q_jh^{(-j)}$
is meaningful in the space of formal Laurent series and
sets up a 1-1 relation between the coefficients $h_i$ and the
standard KP variables $q_i$ of Eq.~\rref{5.1}.
Following ~\cite{Wil81}
one sets $-(Q^j)_-=\sum_{l\ge 1} H_l^jQ^{-l}$
and constructs the Laurent series
\begin{equation}\label{laurent}
\H{n}=z^n+\sum_{l\ge 1}H_l^n z^{-l}.
\end{equation}
The second equation of~\rref{linpb} becomes
\begin{equation}
\dpt{}{n}\psi=\H{n}\psi.
\label{5.9}
\end{equation}
Setting $h\equiv\H{1}$ and $t_1=x$, one gets
$\h{k}=(\del_x^k\psi)/\psi$ and
the obvious continuity equations
$\dpt{}{j} h = \del_x\H{j}$.\\
The facts from  KP theory
which are relevant for us are the following:\\
{\bf a)} The KP equations are equivalent to the conservation laws:
\beq\label{KP}
\dpt{}{n} h =\del_x \H{n},
\eeq
with $H^{(n)}$ of the form~\rref{laurent}.\\
{\bf b)} $\H{k}$ can be expanded as a finite \Fdb\ ``polynomial'':
\beq\label{fdbex}
\H{k}=\h{k}+\sum_{l=0}^{k-1} c_l^k \h{l},
\eeq
with $c_l^k $ independent of $z$.\\
{\bf c)} The $n$--Gel'fand--Dickey reductions of the KP theory
can be defined as the restrictions of the
flows~\rref{KP} to the invariant submanifolds defined by
\begin{equation}
\H{n}=z^n.
\end{equation}
{\bf d)} The key property
linking to the Grassmannian picture~\cite{DBook,SS,sw}
is eq.~\rref{fdbex}.
Let ${\cal H}=L^2(S^1, {\Bbb C})={\cal H}_+\oplus{\cal H}_-$, where
$S^1=\{z\in{\Bbb C}:|z|=1\}$,
${\cal H}_+=\overline{span} \{z^j:j\ge 0\}$,
${\cal H}_-=\overline{span} \{z^j:j<0\}$,
where the bar means $L^2$ closure.
Then the universal Grassmannian~\cite{sw} $Gr({\cal H})$ is the set of
subspaces $W\subset{\cal H}$ for which the orthogonal projections
$\pi_+:W\to{\cal H}_+$ and $\pi_-:W\to{\cal H}_-$
are respectively Fredholm and  Hilbert-Schmidt operators.
The virtual dimension of $W$ is by definition the index of $\pi_+$ and
$Gr({\cal H})$ is the union of connected components labeled by
the virtual dimension.
If $h=z+\sum_{l\ge 1}h_l/z^l$ is a smooth function of $z\in S^1$,
the Fa\`a di Bruno recurrence relations~\rref{recrel}
give a point $W\in Gr$ by
\beq
\label{ww}
W=\overline{span}\{ 1,\h{1},\h{2},\ldots\}.
\eeq
It is not difficult to show that
$\H{k}=\pi_+^{-1}(z^k)$.

\section{Cohomology and the Fa\`a di Bruno recursion relation}
\label{sec3}

Recall how the Krichever map associates a point in the universal
Grassmannian to the datum of a smooth algebraic curve $C$,
a point $p\in C$, a local coordinate $z^{-1}$ centered at $p$, a line
bundle $\cal N$ over $C$ and a local trivialization $\phi_0$ of $\cal N$
in a neighbourhood $U_0$ of
$p$. The local coordinate $z^{-1}$
identifies $S^1=\{z:|z|=1\}$ as a submanifold of $C$, while the sections
of $\cal N$ correspond to functions on $S^1$ via the local trivialization
$\phi_0$.
The point $W\in Gr({\cal H})$ associated to $(C,p,z,{\cal N},\phi_0)$
is the closure in $\cal H$ of the space of sections of $\cal N$ which are
holomorphic on $U_1=C-\{p\}$.
Using the Mayer-Vietoris sequence one shows~\cite{ADKP87,sw}
that the virtual dimension of $W$ is $\chi({\cal N})-1$,
$\chi({\cal N})$ being the Euler
characteristic of $\cal N$.

For completeness, we recall in the following two lemmas some standard facts.
As usual, we will denote by $h^i(C,\CN)=\mbox{dim } H^i(C,\CN).$
\begin{lem}
Let $p\in C$ be a non--Weierstrass point and $z^{-1}$ a local coordinate
vanishing at $p$.
Then the classes of $\{z,z^2,\cdots,z^g\}$ give a basis for
$H^1(C,{\cal O})$ while for any $k>g$ there exists a function $\lambda_k$
on $C$ with a pole of order $k$ at $p$ and without subleading poles
but for those in the Weierstrass gap.\end{lem}

\pf
Let $j,l\ge 0$ and consider the exact sequence
$0\to{\cal O}(jp)\to{\cal O}(lp)\to{\cal S}_{l-j}(p)\to 0$ where
${\cal S}_{l-j}(p)$ is the skyscraper sheaf of length $l-j$ at $p$.
The corresponding cohomology sequence reads
$$0\to H^0(C,{\cal O}(jp))\to H^0(C,{\cal O}(lp))
\to{\Bbb C}^{l-j}{\buildrel{\Psi_4}\over{\to}}H^1(C,{\cal O}(jp))
\to H^1(C,{\cal O}(lp))\to 0.$$
One can use the covering $(U_0,U_1)$ to compute
cohomology.
Since $p$ is not Weierstrass and $h^1(C,{\cal O})=g$, setting $j=0$ and
$l=g$
we see that $H^1(C,{\cal O}(gp))=0$. The classes $[z^k]:=\Psi_4(z^k)$
for $k=1,\cdots,g$ are then a basis of $H^1(C,{\cal O})$.
For every $k>g$, setting $j=k-1$, $l=k$ we obtain that
$H^1(C,{\cal O}(kp))=0$. Then there is a function $\lambda_k$
on $C$ with a pole of order $k$ at $p$.
Moreover, this function is defined up to  sections in $H^0(C,{\cal
O}(k-1)p)$ (and up to homotheties) and
we can use these ambiguities to fix the polar part of
$\lambda_k$ as claimed.
\epf

{}From now on, when we say that a function $f$ has a "simple
pole of order $k$" at $p$ we mean that its Laurent expansion at $p$
is the one stated in the lemma above.

\begin{lem}\label{deform} Assume that $h^0(C,{\cal N})>0$, $h^1(C,{\cal
N})=0$ and let $s\in H^0(C,{\cal N})$ be a nontrivial section.
For all $[\beta ]\in H^1(C,{\cal O})$ and for every  $1-$cocycle $\beta$
representing it there exists an infinitesimal deformation of the couple
$({\cal N},s)$ along $\beta$.
\end{lem}

\pf
Consider the double complex~\rref{complex}.
On the covering $(U_0,U_1)$ we denote by
$\beta_{10}$ a $1-$cocycle representing $[\beta ]$ with a pole of order
$j$ at $p$, by $g_{10}$ the transition function of $\cal N$ w.r.t. some
trivialization and we let $(f_0,f_1)$ be the couple of functions
which represent $s$.
It is clear that $s\beta_{10}$ is a $1-$cocycle and, since
$H^1(C,{\cal N})=0$, it is actually a coboundary.
Hence, there are $0-$cochains
$(\delta f_0,\delta f_1)$ such that
$g_{10}^{-1}\delta f_1=s\beta_{10}+\delta f_0$.
Choosing $\delta g_{10}:=g_{10}\beta_{10}$, we have that
\begin{equation}\label{newsect}
g_{10}^{-1}\delta f_1=\delta f_0+(g_{10}^{-1}\delta g_{10})f_0.
\end{equation}
The couple
$(\tilde f_0,\tilde f_1):=
(\delta f_0+(g_{10}^{-1}\delta g_{10})f_0,\delta f_1)$
is a section of $\cal N$ with a pole of order $j$ at $p$, therefore it
is a holomorphic section of ${\cal N}(jp)$.
\epf

Eq.~\rref{newsect} looks like a pointwise version of the Fa\`a di
Bruno recursion relation with $h=g_{10}^{-1}\delta g_{10}$.
Indeed, this formal resemblance was the starting point of our work.
To accommodate the $x$-dependence of eq.~\rref{recrel}
we need a family version of the construction above.
Actually, the "universal version" is already at hand~\cite{GH}.
We choose in the Krichever data a line bundle ${\cal N}={\cal L}_D$ of
degree $g$ corresponding to a non special effective divisor $D$, in
such a way that $h^0(C,{\cal L}_D)=1$. In other words, there is a
unique (up to homotheties) non trivial section
$\sigma_D\in H^0(C,{\cal L}_D)$ vanishing at $D$.
If $p\in C$ is not in the support of $D$, $\sigma_D$
does not vanish at $p$ and
can be used to trivialize ${\cal L}_D$ in a neighborhood $U_0$ of $p$.
We get in this way the last piece of the
Krichever data up to a ${\Bbb C}^\times$ action.
We shall assume that $p$ is not a Weierstrass point.

Let $C^{(d)}$ be the $d$-th symmetric product of $C$,
whose points are effective divisors $D=\sum_{i=1}^d q_i$.
Recall~\cite{ACGH} that on $Y=C\times C^{(d)}$
there is the universal divisor $\Delta$ of degree $d$,
whose restriction to $C\times \{D\}$ is the divisor $D\subset C$ itself,
and the corresponding line bundle ${\cal O}(\Delta )$.
Let $\mu:C^{(d)}\to {\cal J}(C)$ be
the Abel sum map $\mu (D)=\sum_i\int_{\tilde q}^{q_i}\vec\omega$
with base point $\tilde q\in C$ and $\vec\omega=(\omega_1,\cdots,\omega_g)$
a basis of Abelian differentials on $C$.
We also fix a symplectic basis
$\{a_1,\cdots,a_g,b_1,\cdots,b_g\}$ of $H_1(C,{\Bbb Z})$, normalize
$\vec\omega$ by $\oint_{a_i}\omega_j=\delta_{ij}$ and denote by
$Z_{ij}=\oint_{b_i}\omega_j$ the corresponding period matrix.

The map
$$\tilde{\mu}:C\times C^{(g)}\to {\cal J}(C)$$
defined by $\tilde{\mu}=\mu\circ\pi_1-\mu\circ\pi_2$ ($\pi_i$,
$i=1,2$, being the projection on the $i-$th factor of $C\times C^{(g)}$)
pulls back the theta bundle, translated by the Riemann constant $k$,
to a line bundle ${\cal L}$ on $C\times C^{(g)}$ together with the section
$$\sigma (q,D)=\theta (\mu(q)-\mu(D)-k),$$
$\theta$ being the Riemann's theta function.
The restriction ${\cal L}|_{C\times \{ D\} }$ is a line
bundle of degree $g$ on $C$ with a holomorphic section
$\sigma_D(q):=\sigma (q,D)$.
Notice that $\sigma_D$ is actually an entire function on the
universal covering $\tilde C$ of $C$, which transforms as
$$\sigma_D(q+\vec n\cdot \vec a+\vec m\cdot \vec b)=\sigma_D(q)
\exp(-\pi i\vec m Z\vec m-2\pi i\vec m\cdot(\mu (q)-\mu (D)-k)),$$
under the action of the fundamental group of $C$ generated by $\vec
a=(a_1,\cdots,a_g)$ and $\vec b=(b_1,\cdots,b_g)$.
Since $D$ non special, $\sigma_D$ vanishes precisely at $D$
and hence the above restriction of ${\cal L}$
is isomorphic to ${\cal L}_D$. We denote by
$C_0^{(g)}\subset C^{(g)}$ the open subvariety given by
non special effective divisors, and by $\Delta_0$ the restriction of
the universal divisor $\Delta$ to $C\times C_0^{(g)}$. Summing up

\begin{prop}
The line bundle ${\cal L}$ on $C\times C_0^{(g)}$
is isomorphic to ${\cal O}(\Delta_0)$ and
the section $\sigma$ has divisor $\Delta_0$.\epf
\end{prop}

Denote by $P$ the divisor $P=\{ p\}\times C_0^{(g)}\subset C\times
C^{(g)}_0$. Holomorphic sections $\sigma^{(k)}$ of ${\cal L}((k-1)P)$
are the same as sections of $\cal L$ with poles of order
bounded by $k-1$ along $P$. Their pointwise existence is
obvious, because ${\cal L}((k-1)P)|_{C\times \{D\}}
={\cal L}_D((k-1)p)$ and this last bundle has $k$
holomorphic sections. We shall later construct
a section $\sigma^{(k)}$ for each $k\ge 1$ by means of $\sigma^{(0)}:=\sigma$

\medskip
To link to cohomology, we work out an explicit coordinate description of
$\cal L$ on a subvariety ${\cal B}\subset C\times C_0^{(g)}$ described below.

\noindent
0) Let $U_0$ be an open neighborhood of $p\in C$ with closure
$\bar U_0$ and denote by $C_{0,p}^{(g)}$ the open subset of
$C^{(g)}$ of (non special) effective divisors
$D$ with support not intersecting $\bar U_0$.
We shall work on ${\cal B}=C\times C_{0,p}^{(g)}$.
For every $D\in C_{0,p}^{(g)}$, $\sigma_D$
has divisor $D$ and it does not vanish at $p$.
Therefore, $\sigma$ gives a trivialization of
${\cal L}$ on ${\cal B}_0:=U_0\times C_{0,p}^{(g)}$,
which will be denoted by $\Phi_0$.
Possibly after shrinking, we can assume that
$U_0$ is the domain of a local coordinate $z^{-1}$ centered at $p$.

\noindent
1) A local trivialization $\Phi_1$ of
$\cal L$ on ${\cal B}_1=U_1\times C_{0,p}^{(g)}$, with $U_1=C-\{ p\}$
is constructed as follows. The datum of $p$
gives us the line bundle ${\cal O}(gp)$ on $C$,
together with a holomorphic section
$\sigma_{gp} = \theta (\mu(q)-\mu(gp)-k)$ vanishing of order
$g$ at $p$ and nowhere else.
As a function on the universal covering
$\tilde C$, $\sigma_{gp}$ transforms as
$$\sigma_{gp}(q+\vec n\cdot \vec a+\vec m\cdot \vec b)=\sigma_{gp}(q)
\exp(-\pi i\vec m Z\vec m-2\pi i\vec m\cdot(\mu (q)-\mu (gp)-k)).$$
We look~\cite{Dub,Kr} for a function
$\nu$ on $\tilde C\times C^{(g)}_{0,p}$ such that $\nu\sigma_{gp}$
transforms as $\sigma_D$ and does not vanish on ${\cal B}_1$.
Let $\Omega^{(k)}$, $k>0$, be the unique Abelian differential of
the second kind on $C$ with vanishing $a-$periods and
$w^{-k-1}dw$  as principal part at $p$ ($w=z^{-1}$).
The $b-$periods of $\Omega^{(k)}$ are~\cite{Dub}
$$\Pi^k_l:=\oint_{b_l}\Omega^{(k)}=
{{2\pi i}\over{k!}}{{d^{k-1}\zeta_l}\over{dw^{k-1}}}(p),$$
where $\omega_l(w)=\zeta_l(w)dw,\; (l=1,\cdots ,g)$ is the local form
of the Abelian differentials on $U_0$. Then

\begin{lem}\label{trivial}
For every $D\in C^{(g)}_{0,p}$ there exists an Abelian differential
$\Omega(D)$ of the second kind on $C$,
holomorphically depending on $D$, such that
$\Phi_1=\sigma_{gp}\exp(\int_{\tilde q}^q\Omega(D))$
is a never vanishing section of $\cal L$ on ${\cal B}_1$.
\end{lem}

\pf
$\Phi_1$ has the desired property if and only if
the $a-$periods of $\Omega(D)$ vanish and the $b-$periods are
$\oint_{b_l}\Omega(D)=2\pi i\mu_l(D-gp)=:a_l(D)$,
where $\mu_l$ is the $l-$th component of the Abel map.
If $\Omega(D)=\sum_{k=1}^gb_k(D)\Omega^{(k)}$,
the equation above becomes $$\sum_{k=1}^gb_k(D)\Pi^k_l=a_l(D).$$
Being $p$ not Weierstrass, the matrix $\Pi=(\Pi^k_l)$ is invertible
and the solution for $b_k(D)$ is holomorphic in $D$.
\epf

\begin{rem}
The proof uniquely defines $\Phi_1$. Of course
there are other trivializations,
given by multiplying $\Phi_1$ by a nowhere vanishing
holomorphic function on ${\cal B}_1$, i.e. by adding to $\Omega(D)$ an Abelian
differential of the second kind $\tilde{\Omega}(D)$
with vanishing periods and holomorphic on ${\cal B}_1$.
\end{rem}

Summing up, we have that

\begin{lem}\label{section} In the trivialization $\Phi_i\; (i=0,1)$ given
above, the section $\sigma$ of $\cal L\to B$ corresponds
to the couple $(1,f_1)$ of holomorphic functions on
${\cal B}_0, {\cal B}_1$, where
$$f_1(q,D)={{\theta (\mu(q)-\mu(D)-k)}\over
{\theta (\mu(q)-\mu(gp)-k)}}\exp(-\int_{\tilde q}^q\Omega(D)),$$
and hence $\cal L$ has transition function
$g_{10}=f_1$ on ${\cal B}_1\cap {\cal B}_0$.\epf
\end{lem}

We are now in the position of applying the computations of Appendix~\ref{app}.
We can consider the line bundle $\cal L\to B$ as a deformation of
${\cal L}_{D}$ for every $D\in C^{(g)}_{0,p}$ and the section $\sigma$
as a deformation of $\sigma_D$.
The logarithmic derivative with respect to $D$  of the transition
function
$g_{10}$ is a cocycle representing the cohomology class in $H^1(C,{\cal O})$
of the deformation of the line bundle ${\cal L}_{D}$.

We first consider one dimensional deformations.
Let $\xi :{\cal X}:=C\times X\to {\cal B}$ be 
the identity on $C$ and an embedding of
a disk $X=\{ x\in {\Bbb C}:|x|<\epsilon\}$ into $C^{(g)}_{0,p}$.
For simplicity we will leave implicit the pull-back maps
associated to $\xi$. Set ${\cal X}_i=U_i\times X\; (i=0,1)$ and
denote by $D_x$ the image of $x\in X$ on $C^{(g)}_{0,p}$,
by ${\cal L}_x$ the corresponding line bundle and by $\sigma_x$ its section.

\begin{prop}\label{prop}
There exist

\noindent 1) an embedding of the disk $X$ into $C^{(g)}_{0,p}$,

\noindent 2) a trivialization $\Phi_1$ of $\cal L$ over ${\cal X}_1$,

\noindent 3) sections $\sigma^{(k)}=(f_0^{(k)},f_1^{(k)})$ of
$\cal L$, with a simple pole of order $k\ge 0$ at $P$,

\medskip

\noindent such that, for every $x\in X$,

\medskip

\noindent a) the cohomology class of the deformation of
${\cal L}_x$ is $[\alpha ]$ with representing cocycle $\alpha =
z+\sum_{l>0}\alpha_lz^{-l}$

\noindent b) the cocycle condition~\rref{newsect} for the deformation
$\sigma^{(k)}$ of $\sigma^{(k)}_x$ is the Fa\`a di Bruno recursion
relation~\rref{recrel} with $h=h^{(1)}=\alpha$ and $h^{(k)}=f_0^{(k)}$.
\end{prop}

\pf
Fix a divisor $D_0\in C^{(g)}_{0,p}$. Lemma~\ref{trivial} gives us a
differential $\Omega (D_0)$ and a trivialization of ${\cal L}_0$ on $U_1$.
If $2\pi i\Pi^{(1)}\in {\Bbb C}^g$ are the $b$-periods of $\Omega^{(1)}$,
there is $\epsilon >0$ such that, for $|x|<\epsilon$, $D_x=D_0+x\Pi^{(1)}$
is again in $C^{(g)}_{0,p}$.
We choose for the embedding 1) the map $x\mapsto D_x$ and for the
trivialization $\Phi_1$ the section of
Lemma~\ref{trivial} with $\Omega (D_x)=\Omega (D_0)+x\Omega ^{(1)}$. Then
the transition function of ${\cal L}_x$ is
$$g_{10}(q,x)={{\theta (\mu(q)-\mu(D_0)-x\Pi^{(1)}-k)}\over
{\theta (\mu(q)-\mu(gp)-k)}}\exp(-\int_{\tilde q}^q\Omega(D_x))$$
and $\alpha=g_{10}^{-1}\partial_xg_{10}$ satisfies a).
As for 3), the section $\sigma^{(0)}=\sigma$ of Lemma~\ref{section}
is a deformation of $\sigma_x$ and the corresponding cocycle
condition reads
$$g_{10}^{-1}\partial_xf_1^{(0)}=\partial_xf_0^{(0)}+\alpha f^{(0)}_0,$$
showing that the couple
$(f^{(1)}_0,f^{(1)}_1):=(\partial_xf^{(0)}_0+\alpha
f^{(0)}_0,\partial_xf^{(0)}_1)$
is a holomorphic section $\sigma^{(1)}_x$ of ${\cal L}_x(p)$. Since
$\sigma^{(1)}_x$ depends holomorphically on $x$, we have a section
$\sigma^{(1)}$ of ${\cal L}(P)$. Iterating this procedure one constructs
all the other sections $\sigma^{(k)}$ for $k>1$.
Setting $h^{(k)}:=f_0^{(k)}$ one has that $h=h^{(1)}=\alpha$ and the
cocycle condition above for the deformation $\sigma^{(k)}$ of
$\sigma^{(k)}_x$ gives precisely the Fa\`a di Bruno recursion
relations.
\epf

\begin{rem}
There is a simple connection between our construction and the
Baker-Akhiezer function of~\cite{Dub,Kr,sw}.
Indeed, $g_{10}(z,x)=f_1^{(0)}(z,x)|_{U_1\cap U_0}$, and $f_1^{(0)}(z,x)$ is a
holomorphic function on $U_1$ whose zeros define the effective divisor
$D_x$ corresponding to the line bundle ${\cal L}_x$.
Now, $\psi(z,x):=f_1^{(0)}(z,x)/f_1^{(0)}(z,0)$ is meromorphic on $U_1$,
its poles correspond to the non-special effective divisor $D_0$
of degree $g$, and it has an essential singularity at $p$,
i.e. it is a Baker-Akhiezer function.
This gives another justification to our definition of $h$ as
$\partial_x\log f_1^{(0)}(z,x)=\partial_x\log\psi(z,x)$.
\end{rem}

\medskip

The full KP hierarchy has a similar cohomological meaning, which will be
quickly sketched below. More details will be given elsewhere.
Let us go back to the universal family ${\cal L}\to {\cal B}$. If
$t_1,\cdots ,t_g$ are local coordinates on $C^{(g)}_{0,p}$,
the classes of $g_{10}^{-1}\partial_{t_k}g_{10}$,
$k=1,\cdots,g$ are a basis of $H^1(C,{\cal O})$ and give vector fields on
$C^{(g)}_{0,p}$. We can choose $t_1=x$.
For every $j>g$ we introduce an extra parameter $t_j$ and change the
transition function to
$$\tilde{g}_{10}(z,\vec t)=g_{10}(z,t_1,\cdots,t_g)e^{\sum_jt_j\lambda_j(z)},$$
where $\lambda_j(z)$ is a meromorphic function on $C$ with a simple pole of
order $j$ at $p$ and holomorphic elsewhere.
The new transition function belongs to the same cohomology class of
the old one, so the family of line bundles ${\cal L}$ is unaffected:
we have only changed the trivialization over ${\cal B}_1$.
As a result, the image of the Krichever map~\rref{ww} is the same as before.
The motion of the point $W_{\vec t}$ in the Grassmannian is easily described
as follows. Let $\bar \Gamma (U_1,{\cal O})$ be the closure in
$\cal H$ of the span of the holomorphic functions
on $U_1$. Then
 \begin{equation}\label{motion}
W_{\vec t}=g^{-1}_{10}(z,\vec t)\bar
\Gamma (U_1,{\cal O})={{g_{10}(z,0)}\over{g_{10}(z,\vec t)}}W_0.
\end{equation}
Notice that $W_{\vec t}$ does not depend on the choice for the
trivialization of ${\cal L}_{\vec t}$ over $U_1$.

We apply the construction of proposition ~\ref{prop} to the section $\sigma
=(1,f_1)$
along all the vector fields represented by
$\alpha_k=g_{10}^{-1}\partial_kg_{10}$ getting a cocycle condition of
the form
$$H^{(k,j+1)}:=g_{10}^{-1}\partial_k^{j+1}f_1=
\partial_kH^{(k,j)}+\alpha_kH^{(k,j)}$$
where $\partial_k=\partial/\partial t_k$ and the tildes have been dropped.
It is clear that $H^{(k,j)}\in W_{\vec t}$ (for all
$k,j\in{\Bbb N}$) and, by construction, $\{H^{(k)}:=H^{(k,1)}:k\in{\Bbb N}\}$
is a basis of $W_{\vec t}$. To get this last identification one chooses
the coordinates $t_1,\cdots ,t_g$ in such a way that the $\alpha_k$ have
a simple pole of order $k$ at $p$.
Differentiating with respect to $t_j$ the identity
$\partial_kf_1=g_{10}H^{(k)}$
(and observing that the right hand side times $g_{10}^{-1}$ is an element
of $W_{\vec t}$ and can therefore be expressed as a linear combination
of the basis elements $H^{(l)}$) one gets that the equation of motion
{}~\rref{motion} is equivalent to the set of differential equations
\begin{equation}\label{hypcohKP}
\left(\partial_j+H^{(j)}\right)H^{(k)}=\sum_{l\in{\Bbb N}}c^{jk}_lH^{(l)},
\end{equation}
found in ~\cite{cfmp}.
Finally, the Fa\`{a} di Bruno basis reads
$\{h^{(j)}:=H^{(1,j)}:j\in{\Bbb N}\}$ and ~\rref{hypcohKP} is equivalent to
the conservation laws ~\rref{KP}.

\section{An example: elliptic curves}
\label{sec4}

The simplest example is when $C$ is
an elliptic curve, which we identify with its own Jacobian.
Fix $p$ and $\tilde{q}$ in $C$. Then $C^{(1)}_{0,p}=C-\bar U_0-\{k\}$
where $k$ is the Riemann constant.
Thus, $\Omega^{(1)}(w)=(\wp (w)+c)dw$ where $w$ is a uniformizing coordinate
on $C$ centered at $p$, $\wp(x)$ is the Weierstrass function and
$c=-\oint_a\wp(w)dw$.
Since $\wp(w)=-{{d^2}\over{dw^2}}\log\theta_{11}(w)$
up to a constant, it follows that
$$\int_{\tilde{q}}^{q(w)}\Omega^{(1)}(w)=
-{{\theta_{11}^\prime(w)}\over{\theta_{11}(w)}}$$
up to a constant which we can neglect.
The $b-$period of $\Omega^{(1)}$ is $2\pi i$ and we find
$$g_{10}(z,x)=f^{(0)}_1(z,x)=
{{\theta(z^{-1}-q_1-x-k)}\over{\theta(z^{-1}-k)}}\exp\left(-(q_1+x)
{{\theta_{11}^\prime(z^{-1})}\over{\theta_{11}(z^{-1})}}\right),$$
where $z=w^{-1}$, $D_0=q_0$ and $D_x=q_0+x$.
The Baker-Akhiezer function reads
$$\psi(z,x)={{\theta(z^{-1}-q_1-x-k)}\over{\theta(z^{-1}-q_1-k)}}
\exp\left(-x{{\theta_{11}^\prime(z^{-1})}\over{\theta_{11}(z^{-1})}}\right),$$
and the Fa\`a di Bruno generating function is
$$h(z,x)=-{{\theta^\prime(z^{-1}-q_1-x-k)}\over{\theta(z^{-1}-q_1-x-k)}}-
{{\theta_{11}^\prime(z^{-1})}\over{\theta_{11}(z^{-1})}}.$$
The other elements of the Fa\`a di Bruno basis are obtained
recursively, the functions $H^{(k)}$ for $k>1$ can be written in terms of the
$\wp-$function and its derivatives and the flows generated by $t_k$
are trivial.

Finally, we want to explain how one can recover the Jacobian of the curve
$C$ from the Fa\`a di Bruno polynomials. Since the only non--trivial flow
in the KP equations~\rref{KP}
is the first, we can restrict $h$ to
satisfy
\begin{equation}\label{s2}
H^{(2)}\equiv \h{2}-2 h_1 =z^2
\end{equation}
\begin{equation}\label{fl3}
\partial_xH^{(k)}=0\quad \forall k\ge 3
\end{equation}
The first
condition~\rref{s2}
allows to represent the $x-$derivatives of the Laurent coefficients
$h_k$ of $h$ as ordinary polynomials in the same coefficients:
$h_{kx}=P_k(h_1,\cdots,h_{k+1})$, e.g.
$$\begin{array}{lll}
h_{1x}=-2h_2, & & h_{2x}=-2h_3-h_1^2, \\
h_{3x}=-2h_4-2h_1h_2, & & h_{4x}=-2h_5-2h_1h_3-h_2^2,
\end{array}$$
and the same is valid for the Laurent coefficients of $H^{(k)}$.
For $k=3$ we have
$$
H^{(3)}_{-1}=h_3-h_1^2, \quad H^{(3)}_{-2}=h_4-h_1h_2, \quad
H^{(3)}_{-3}=h_5-h_1h_3.
$$
Using the condition~\rref{fl3} for $k=3$, we infer that $h_3=c_1+h_1^2$
and $h_5=c_3+h_1h_3$, where $c_1$ and $c_3$ are the {\em constants}
$c_1=H^{(3)}_{-1}$, $c_3=H^{(3)}_{-3}$.
Thus we obtain $h_5=h_1^3+c_1h_1+c_3$
which, by means of the previous relations, takes the form
$$h_{1x}^2=4 h_1^3+8c_1h_1+8 c_3.$$
We see immediately that this is the Weierstrass equation for the
$\wp-$function after the identifications $h_1(x)=\wp(x)$, $8c_1=-g_2$
and $8c_3=-g_3$, so that we have an elliptic curve with uniformizing
coordinate $x$.
{}From our discussions it is clear that this is the Jacobian
${\cal J}(C)\cong C$ of $C$ itself, since $x$ is the parameter for the
deformations of $\cal L$.
This obviously reflects the well known result~\cite{Dub,sw} which
expresses the solution $u=2h_1$ of the KdV equation
as the second logarithmic derivative of the
theta function of ${\cal J}(C)$.

\appendix
\section{Basic facts on deformation theory}\label{app}

We collect here~\cite{Hitchin,W} some notions of
Kodaira--Spencer deformation theory used in the paper.
Let $C$ be a smooth algebraic curve, $\cal N$ a line bundle over $C$ with
a non trivial holomorphic section $s$.
A deformation of the couple $({\cal N}, s)$, with parameter space a ball
${\cal B}\subset{\Bbb C}^n$, is a couple $({\cal L}, \sigma)$
where
\begin{description}
\item[a)] $\cal L\to C\times{\cal B}$ is a line bundle together with an
isomorphism between $\cal N$ and ${\cal L}_0:={\cal L}|_{C\times\{0\}}$,
\item[b)] $\sigma$ is a holomorphic section of $\cal L$ such that
$\sigma_0:=\sigma|_{C\times\{0\}}=s$.
\end{description}
We can cover $C\times{\cal B}$ with open subsets of the form
${\cal U}_j:=U_j\times{\cal B}$ (with local coordinates $z_j$
on $U_j$ and $t=(t_1,\cdots,t_n)$ on $\cal B$) over which $\cal L$
trivializes with fibre coordinate $\xi_j\in{\Bbb C}$.
On the overlaps there exist transition functions $g_{jk}(z_k,t)$
such that $\xi_j=g_{jk}(z_k,t)\xi_k$ and satisfying the cocycle condition
$$g_{jk}(t)g_{kl}(t)=g_{jl}(t).$$
One can assume that
$g_{jk}(t=0)$ are the transition functions of $\cal N$.
The section $\sigma$ of $\cal L$ is given by local functions
$f_j(z_j,t)$ on ${\cal U}_j$ which glue as
$$f_j(z_j,t)=g_{jk}(z_k,t)f_k(z_k,t).$$
The functions $f_j(t=0)$ represent the section $s$.
The infinitesimal version of the relations above at $t=0$ reads
\beq\label{cocy}
g_{jk}^{-1}\partial_tg_{jk}+g_{kl}^{-1}\partial_tg_{kl}-
g_{jl}^{-1}\partial_tg_{jl}=0,
\eeq
showing that $g_{jk}^{-1}\partial_tg_{jk}$ is a
$1-$cocycle with values in $\cal O$, and
\beq\label{sect}
\partial_tf_j-g_{jk}\partial_tf_k=\left(\partial_tg_{jk}\right)f_k.
\eeq
Changing the transition functions by a coboundary
($\tilde{g}_{jk}=g_jg_{jk}g_k^{-1}$) adds to the $1-$cocycle above the
coboundary $g_j^{-1}\partial_tg_j-g_k^{-1}\partial_tg_k$.
Accordingly, the isomorphism classes of infinitesimal deformations of
$\cal N$ correspond to the elements of $H^1(C,{\cal O})$.
Of course the action of the $1-$coboundaries of $\cal O$ extends to the
infinitesimal deformation of $s$ by mapping $\partial_tf_j$ to
$\partial_tf_j+g_j^{-1}\partial_tg_j$.

All this information can be collected in the following hypercohomology
group. Consider the sequence
$$0\to{\cal O}{\buildrel{s\cdot}\over{\to}}{\cal N}\to 0,$$
where $s\cdot$ is the multiplication by $s$, and the double complex
given by taking \v{C}ech cochains
\begin{equation}\label{complex}
\begin{array}{ccccc}
{\cal C}^0(C,{\cal O}) & {\buildrel{\delta}\over{\to}} &
     {\cal C}^1(C,{\cal O}) & \to & \cdots \\
\downarrow s\cdot &  & \downarrow s\cdot & & \\
{\cal C}^0(C,{\cal N}) & {\buildrel{\delta}\over{\to}} &
{\cal C}^1(C,{\cal N}) & \to & \cdots
\end{array}
\end{equation}
where $\delta$ is the coboundary operator.
Set ${\cal A}^p:={\cal C}^p(C,{\cal O})\oplus{\cal C}^{p-1}(C,{\cal N})$
and define the operator $\delta_s:{\cal A}^p\to{\cal A}^{p+1}$ by
$\delta_s(u,v)=(\delta u,\delta v+(-1)^psu)$.
Then $\delta_s^2=0$ and one defines the hypercohomology groups
${\Bbb H}^p_s$ as the cohomology groups of $({\cal A}^\cdot,\delta_s)$.
By ~\rref{cocy} and~\rref{sect},
$\rho:=(g_{jk}^{-1}\partial_tg_{jk},\partial_tf_j)\in{\cal A}^1$
is actually a cocycle in this hypercohomology.
Zero cochains $(g_i,0)\in {\cal A}^0$ give rise to $1$-coboundaries
of the form $(g_i-g_j,sg_i)$ and $\rho+(g_i-g_j,sg_i)$ corresponds
to an isomorphic deformation. Hence,
the isomorphism classes of infinitesimal deformations of
the couple $({\cal N},s)$ correspond to the elements of ${\Bbb H}^1_s$.
\subsection*{Acknowledgements}
We would like to thank F. Magri for discussions and B. Dubrovin
for a careful reading of the manuscript.

\end{document}